\documentclass[epsf,aps,twocolumn,showpacs,preprintnumbers,amsmath,amssymb,footinbib]{revtex4}
\usepackage{epsf}
\usepackage[colorlinks=true, pdfstartview=FitV, linkcolor=red, citecolor=blue, urlcolor=blue]{hyperref}
\newcommand{\beq}{\begin{equation}}

\def\ibid#1#2#3{{\it ibid.} {\bf #1}, #2 (#3)}

\def\npa#1#2#3{Nucl. Phys. A {\bf #1}, #2 (#3)}
\def\npb#1#2#3{Nucl. Phys. B {\bf #1}, #2 (#3)}

\def\prd#1#2#3{Phys. Rev. D {\bf #1}, #2 (#3)}

\def\Tc{T_\text{c}}
\def\Tqk{T_\text{qk}}
\def\Nc{N_\text{c}}
\def\Nf{N_\text{f}}
\def\muQ{\mu_\text{Q}}
\def\muB{\mu_\text{B}}
\def\SU{SU}

\begin{document}
\title{Baryons and the phase diagram for a large number of colors and flavors}
\author{Yoshimasa Hidaka,$^{a}$
Larry D. McLerran,$^{a,b}$ and Robert D. Pisarski$^{b}$}
\affiliation{
$^a$RIKEN BNL Research Center, Brookhaven National Laboratory, Upton, NY 11973, USA\\
$^b$Department of Physics, Brookhaven National Laboratory, Upton, NY 11973, USA\\
}
\begin{abstract}
We consider the possible phases of a non-Abelian gauge theory, 
as a function of temperature and quark chemical potential,
when both the number of colors and flavors is very large.
Generally, a large number of flavors washes out deconfining phase
transitions.  
We show, however, that the degeneracy of even the lightest baryons
is exponentially large.  This implies that the baryon
number (or fluctuations thereof, at zero chemical potential)
is an order parameter in the limit of an infinite number of colors and flavors.
\end{abstract}
\date{\today}
\maketitle

\section{Introduction}

Understanding the phase transitions of  non-Abelian gauge
theories in various limits may  provided a qualitative 
understanding of the phase transitions of QCD.  One limit is to take 
the number of quark flavors, $\Nf$, fixed as the number of colors, $\Nc$,
becomes large \cite{Hooft}.  Quarks are suppressed in this limit, so the
transition is dominated by gluons; at nonzero temperature, there is
expected to be a deconfining transition of first order.  
It is not obvious what happens as 
one moves out in the plane of quark chemical potential.
In ref.~\cite{mcl}, it was suggested that for low temperatures, there
is a ``quarkyonic'' phase, in which quarks are confined, but dominate the
free energy, up to power like corrections.

In this note we make an elementary comment about the limit where the number
of quark flavors, $\Nf$, is allowed to grow along with $\Nc$
\cite{veneziano}.  In the
first section, we review what a standard analysis indicates.  
The potential
between quarks approaches a constant at large distances, as
quark-antiquark pairs
short out the linear confining force.
Thus there is no order parameter which strictly vanishes in a confined phase:
the theory is always deconfined.
Of course with massless quarks, 
it is possible to have a chiral phase
transition.  For massive quarks, however, there is no obvious order
parameter, 
and naively one concludes that there is no phase transition
in plane of temperature, $T$, and baryon chemical potential, $\muB$.

We argue that the expectation value of the baryon number provides an 
order parameter in the limit in which both $\Nc$ and $\Nf$ are infinite.
We show that for $T$  and quark chemical potential $\muQ = \muB/\Nc$ 
somewhat smaller
than the scale parameter of the theory, $\Lambda_\text{QCD}$, 
there are mesons and glueballs present, but no baryons.
In the plane of
$T$ and $\muQ$, above some line the baryon number 
acquires a non-zero expectation value, and thus there is a phase
transition across this line.
(When $\muQ = 0$, there is no net baryon number, so there
one must use fluctuations in baryon number.)
When baryons condense, there is no way to tell if they (or anything else)
are confined or not.
This phase is therefore analogous to the quarkyonic phase 
of large $\Nc$ and small $\Nf$ \cite{mcl}.  We also explore
the way the phase transition evolves at large $\Nc$, as we allow
$\Nf$ from a small value, $\ll \Nc$, to one $\sim \Nc$.

\section{Expansion in Large $\Nc$ and $\Nf$}

The usual analysis for a large number of colors and flavors is familiar
\cite{veneziano}.  We start by considering the theory
for small $\Nf$, and large $\Nc$, and then consider how things change
when $\Nf$ grows with $\Nc$.

At small $\Nf$, in vacuum
there is a confined phase, in which all states are color singlets.
At low temperature, 
it is an ideal phase at infinite $\Nc$: the trilinear interactions
between mesons is $\sim 1/\sqrt{\Nc}$, while those between glueballs
are $\sim 1/\Nc$; quartic interactions are $\sim 1/\Nc$ and $\sim 1/\Nc^2$,
respectively.  Thus the free energy of the confined phase is $\sim 1$,
up to corrections of $\sim 1/\Nc$.  

At a nonzero temperature $T$, and zero quark chemical potential, we
assume that there is a deconfining transition at a temperature $\Tc$.
Like the masses of the lightest mesons and glueballs, we take this to 
be $\sim 1$ as $\Nc \rightarrow \infty$.  Above $\Tc$, the free energy
from the deconfined gluons is $\sim \Nc^2$, so that the free energy
itself is an order parameter.  

The behavior of the transition in the plane of temperature and quark
chemical potential, $\mu$, is not obvious.  For values of $\mu$ which
are finite as $\Nc \rightarrow \infty$, $\Tc$ is independent of $\mu$.
Once a Fermi sea of quarks condenses, for $\mu$ greater than some mass
threshold, at $T < \Tc$ there is a novel phase, termed
``quarkyonic'' \cite{mcl}: while the free energy is, up to the
power like corrections, that of free quarks, excitations near the Fermi
surface are those of confined states, which are baryons.  

Even so, the pressure provides an order parameter: it is $\sim 1$ in
the confined phase, $\sim \Nc$ in the quarkyonic phase, and $\sim \Nc^2$
in the deconfined phase.  

\begin{figure}
\begin{center}
\epsfxsize=.40\textwidth
\epsfbox{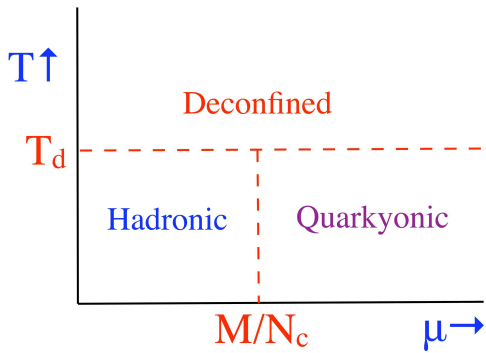}
\end{center}
\caption{Phase diagram for $\Nc = \infty$ and finite $\Nf$.}
\end{figure}

We make one obvious comment which will 
matter latter.  As illustrated in Fig. (1), consider the hadronic
phase, which exists in a ``box'' where $T < \Tc$
and for values of $\mu$ below the mass threshold \cite{mcl}.  
Baryons are composed of
$\Nc$ quarks, and so have a mass $\sim \Nc$.  Simply because they are so heavy,
at infinite $\Nc$ the hadronic phase only contains mesons and glueballs,
and no baryons.  
There are baryons in the quarkyonic and deconfined phases, although in the
deconfined phase, they cannot be distinguished from quarks.

Now consider letting the number of flavors increase along with the number
of colors.  We assume that this is done so that asymptotic freedom still
holds; in practice this is not very restrictive.

Consider the spectrum in vacuum.  States need to be gauge invariant
with respect to color, but can 
carry flavor.  We assume there are $\Nf$ flavors of quarks, but 
for the time being, we do not
need to assume anything about further flavor or chiral symmetries.

Mesons are composed from a quark and an anti-quark, and so for each spin,
there are $\sim \Nf^2$ mesons.   
Up to factors involving functions of $\Nf/\Nc$, the
couplings between mesons and glueballs have the same parametric
dependence upon 
$\Nc$ as they do for $\Nf/\Nc \rightarrow 0$.  
Nevertheless, in loop diagrams all possible species
of mesons contributes, with the sum over flavor canceling the suppression
associated with small couplings at large $\Nc$.  

For example, consider a meson $M_1^{i j}$, 
where $i$ and $j$ are flavor indices.  At one loop order, its 
self energy receives a contribution from processes such as
$M_1^{i j} \rightarrow M_2^{i k} M_3^{k j} 
\rightarrow M_1^{i j}$.  
Each coupling between three mesons is $\sim 1/\sqrt{\Nc}$,
but because of the sum over the flavor index $k$, representing the
$\sim \Nf$ possible intermediate states, in all the self energy
$\sim \Nf/\sqrt{\Nc}^2 \sim \Nf/\Nc$, which is of order one.
Thus while the individual couplings between specific mesons and glueballs
are each small at large $\Nc$, because there are so many
possible mesons to interact with, at large $\Nf$ mesons (and glueballs)
are strongly interacting.  This is very unlike the case of large $\Nc$
and small $\Nf$, where in vacuum, or at low temperature and density,
mesons and glueballs only weakly interact.

It is also clear that the pressure is no longer an order parameter.
At high temperature, and/or large values of the quark chemical potential,
the pressure can be computed perturbatively, 
and receives contributions $\sim \Nc^2$ from gluons,
and $\sim \Nc \Nf$ from quarks. 
If both $\Nf$ and $\Nc$ are large, this
is of the same order as the mesonic contribution to the pressure, which
is $\sim \Nf^2$. The only statement one can make is that because they
don't have a large flavor degeneracy, the contribution of glueballs to
the pressure is irrelevant, only $\sim 1$.

This is reflected in the following.
In a pure gauge theory, there is a strict order parameter for deconfinement,
given by the value of the renormalized Polyakov loop.  
At small $\Nf$, and infinite $\Nc$, the expectation value of the Polyakov
loop vanishes in the confined phase, and approaches one at high temperature.
Quarks contribute to the Polyakov loop as $\sim \Nf/\Nc$, 
and so at large $\Nf$, the renormalized Polyakov loop is nonzero.
The value of the Polyakov 
loop increases with temperature, and may even jump, 
if there is a first order transition.
Even so, it is of order one at all $T$.  This is like the pressure,
which is of the same order in all phases.  

Consequently, unlike small $\Nf$ and large $\Nc$, when 
both $\Nf$ and $\Nc$
are large, nether a Polyakov loop nor the pressure provide an order parameter.
As noted before, there can still be a chiral phase transition associated
with massless quarks.

At nonzero quark chemical potential, baryons also contribute to the free
energy.  As well shall see in the next section, because of the great
multiplicity of the baryons, it is better viewing their contribution to
the free energy as from highly non-ideal quarks, as in the quarkyonic
phase at small $\Nf$ \cite{mcl}

\section{Baryon Degeneracy}

We now assume that the $\Nf$ flavors of quarks respect an exact flavor
symmetry of $SU(\Nf)$, and compute the degeneracy of possible baryons.
We consider the simplest baryons with
spin $j+1/2$, assuming that $j$ is finite as 
$\Nc$ and $\Nf \rightarrow \infty$.
These are the lowest energy states
in spin and flavor; in a non-relativistic quark model
with three flavors, they correspond to the octet and 
decuplet baryon multiplets.
 
\begin{figure}
\begin{center}
\epsfxsize=.40\textwidth
\epsfbox{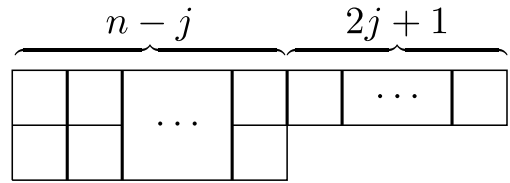}
\end{center}
\caption{Young tableaux corresponding to spin $j+1/2$.}
\label{YoungTableaux}
\end{figure}
 
For convenience, we define 
$$
\Nc = 2 n + 1 \; .
$$  
A baryon is composed of $\Nc$
quarks, so we need to classify the degeneracy with respect to $\SU(\Nf)$.
We assume that the wave function is antisymmetric in color,
so that we need to construct a wave function which is totally
symmetric in spin and flavor.
Since the spin group is $\SU(2)$, 
we can construct a state with spin $j+1/2$ 
by taking a flavor Young tableaux to have at most two rows, 
as in Fig.~(\ref{YoungTableaux}); this is
Fig. (8) of Luty, \cite{veneziano}.
It has $n-j$ boxes in the second row,
and $n- j$ plus $2 j + 1 = n + j + 1$ boxes in the first row.
For the $n-j$ columns with two rows, 
in each column the Young tableaux requires us to antisymmetrize with respect
to flavor.
To obtain a wave function which is symmetric in spin
and flavor, then, for each such pair we also
antisymmetrize with respect to spin.  Hence 
each of the $n-j$ columns contains a pair of quarks with zero spin.
This leaves $2 j + 1$ columns with just one row; the symmetric
combination of these obviously gives a state with spin $j+1/2$.  

The computation of the degeneracy of this Young tableaux is then a simple
application of the ``factors over hooks'' rule for the dimension of a
representation \cite{georgi}.  This gives
\begin{equation}
d_j = \frac{(2 j + 2) \; (\Nf + n + j)! \; (\Nf + n - j - 2)! }{(\Nf - 1)! \; (\Nf - 2)! \; (n + j + 2)! \; (n - j)!} \; .
\end{equation}
The first factor, $2j+2$, is the usual degeneracy of a state with spin $j+1/2$.
The other factors are special to nonzero $\Nc$ and $\Nf$.  These states neglect
the possibility of orbital angular momentum, and so are clearly amongst the
lightest baryons.

It is then direct to compute the degeneracy of this state as $\Nc$ and $\Nf$
become large:
$$
d_j \sim \frac{4 (j + 1)}{\pi \;(\Nc -1 )^2 } \; \frac{\xi^2}{1+\xi}
\; \exp\left( \; (\Nc -1 ) \; f(\xi)  \right)\; ,
$$
where
\begin{equation}
\xi = \frac{2 \Nf}{\Nc - 1} 
\;\;\; , \;\;\;
f(\xi) = (1 + \xi) \log\left(1 + \xi\right) - 
\; \xi  \log\left( \xi \right) \; .
\end{equation}
Since $df(\xi)/d\xi = \log(1 + 1/\xi)$, $f(\xi)$ increases
monotonically, as $\xi$ varies from $0$ to $\infty$.
$f(\xi)$ has a cusp about the origin, 
\begin{equation}
f(\xi) \sim  \xi \left( 1 - \log(\xi)  \right)
\;\;\; , \;\;\; \xi \rightarrow 0 \; ;
\label{smallxi}
\end{equation}
thus there is a smooth limit when $\Nf/\Nc$ vanishes.  For large $\xi$,
$f(\xi)$ grows logarithmically,
\begin{equation}
f(\xi) \sim + \log(\xi) \;\;\; , \;\;\; \xi \rightarrow \infty \; .
\end{equation}

If $\Nf$ is held fixed as $\Nc \rightarrow \infty$, then 
the degeneracy of the lightest baryon is
$\sim \exp( \Nc f(\xi))/\Nc^4 \sim \Nc^{- 4 + 2\Nf}$.  

In contrast, if $\Nf$ is allowed to grow with $\Nc$,
then $\xi\sim 2\Nf/\Nc$ is a number of order one,
and the degeneracy of these baryons --- which are, admittedly, a small
subset of all baryons --- increases exponentially, as $\Nc$ times the
function $f(\xi)$.  
While mathematically trivial, this increase in the degeneracy
of even the lightest baryons has dramatic consequences for thermodynamics.
Even though the baryons are very heavy, and so Boltzmann
suppressed at nonzero temperature, at finite temperature 
this can be overwhelmed by the large degeneracy of states.
In other words, the entropy times temperature, $T\ln d_j$, 
overwhelms the energy, $m_\text{B}$.
If $m_\text{B}$ is the mass of the lightest
baryon, then at vanishing quark chemical potential, this occurs when
the temperature is greater than the ``quarkyonic'' temperature,
\begin{equation}
\Tqk = \frac{m_\text{B}}{\Nc} \; \frac{1}{f(\xi)} \; .
\end{equation}
Above $\Tqk$, while there is no net density of baryons,
fluctuations in baryon number
--- or, equivalently, quark number --- are unsuppressed.
At nonzero quark density, the corresponding temperature is
\begin{equation}
\Tqk(\mu) = \left(\frac{m_\text{B}}{\Nc} - \mu \right) 
\; \frac{1}{f(\xi)} \; .
\label{triangle}
\end{equation}
This vanishes when $\mu = m_\text{B}/\Nc$, which is the naive mass 
threshold for
producing a Fermi sea of baryons at zero temperature.  This mass threshold
can be corrected by effects from binding in nuclear matter \cite{mcl}.

By its nature, $\Tqk$ is well defined {\it only} when both
$\Nc$ and $\Nf$ are infinite.  When $\Nc$ and $\Nf$ are finite, there is a 
large, but finite, degeneracy for the ground state baryons, and 
at any finite temperature, that cannot
win out over Boltzmann suppression.

We do not claim that our ``computation'' of $\Tqk$
is anything more than suggestive.  Baryons interact strongly, and there are
many more excited baryons than we have considered.  It is difficult to
imagine, however, that the effects of interactions, or other baryonic states,
could suppress the exponentially large increase above.  
For example, while meson-baryon and baryon-baryon interactions are large,
they are large by powers of $\Nc$, not exponentials.

\section{The Phase Diagram as a Function of $\Nf/\Nc$}

We conclude with some speculations as to how the phase diagram might
change at infinite $\Nc$, as one tunes the number of flavors from
being finite, to growing with $\Nc$.

The phase diagram when $\Nf/\Nc$ vanishes as $\Nc \rightarrow \infty$
is shown in Fig.~(1).  In this limit, there is no net density of baryons,
or even fluctuations thereof, in the hadronic ``box''.  There are baryons,
or baryonic fluctuations (when $\mu = 0$) in the quarkyonic and deconfined
phases.  Thus we simply ask how the phase diagram of Fig.~(1) could
smoothly interpolate as we increase $\Nf/\Nc$.

At small values of $\Nf/\Nc$, presumably one has the phase diagram
of Fig.~(2).  The hadronic ``box'' for small $T$ and $\mu$ now has
a curved boundary.  Further, the line separating the deconfined and
quarkyonic phases surely persists for some nonzero range in $\Nf/\Nc$.
One can then speak of two phases of the theory, hadronic and quarkyonic.
Clearly the terminology is imprecise: for very small values of $\Nf/\Nc$,
the phase at high temperature, and low $\mu$, will look like a deconfined
phase, dominated by gluons, with some admixture of quarks.  Thus the
persistence of the line separating the deconfined and quarkyonic phases
indicates the quantitative difference between the two phases.

\begin{figure}
\begin{center}
\epsfxsize=.40\textwidth
\epsfbox{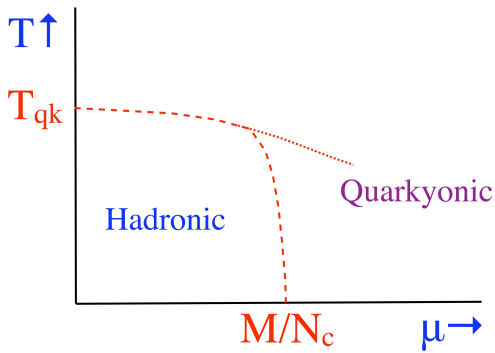}
\end{center}
\caption{Possible phase diagram for $\Nc=\infty$, and 
a small, but nonzero value of $\Nf/\Nc$.}
\end{figure}

Now let $\Nf/\Nc$ continue to increase to large values.  As shown
in Fig. (3),
there is certainly a phase boundary between the hadronic and quarkyonic
phases, but there is no point in speaking of a deconfining transition.
Thus as indicated in the figure, at some point the line between the deconfined
and quarkyonic phases presumably shrinks away.
Further note that unlike (\ref{triangle}), we do not draw the boundary
between the two phases as a straight line.  Due to interactions between
mesons and baryons, we expect that this will acquire some curvature.

\begin{figure}
\begin{center}
\epsfxsize=.40\textwidth
\epsfbox{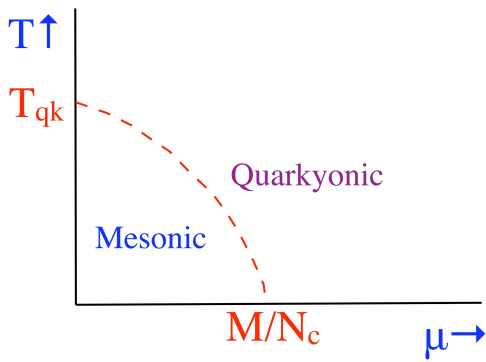}
\end{center}
\caption{Possible phase diagram for $\Nc = \infty$ 
and $\Nf/\Nc$ of order one.}
\end{figure}

We have assumed only a flavor symmetry of $\SU(\Nf)$.
Alternately, assume that there are $\Nf$ flavors of
massless quarks, with a 
$\SU(\Nf)\times \SU(\Nf)$ chiral symmetry which is spontaneously
broken to $\SU(\Nf)$ in vacuum.  
(The axial $U(1)$ symmetry is 
a correction $\sim 1/\Nf^2$ at large $\Nf$, 
and so of little consequence.)
In the chirally broken phase, our arguments go through as before.
At nonzero temperature or density, however, 
there is now another line of phase transition in Figs. (1), (2), and (3),
to restore chiral symmetry.  When
$\Nc = \infty$ and $\Nf/\Nc = 0$ \cite{mcl},
the chiral transition is expected to split from deconfinement at the
corner of the hadronic box, extending into the quarkyonic phase.  This
presumably is true of Fig. (2), where $\Nf/\Nc$ is small but nonzero.
As $\Nf/\Nc$ continues to increase, however, there is no reason for
the chiral transition to continue to coincide for the line characterizing
baryon condensation (or fluctations).  Indeed, since deconfinement is
eventually completely washed away, it is possible that the chiral transition
completely separates from the baryonic/quarkyonic transition line of Fig. (3),
even at $\mu = 0$.  We do not draw all of the possibilities, since we
can only resort to conjecture.

It is impossible to resist the completely unreasonable 
extrapolation from infinite $\Nc$ and $\Nf$ to finite values.
For three colors and two flavors,
$f(2) \approx 1.91...$; for three colors and flavors,
$f(3) \approx 2.25...$.  With $m_\text{baryon}/3 = 313$~MeV,
$\Tqk \approx 160$ and $\approx 140$~MeV, respectively.
At least these values are where we expect something to happen in QCD.

\section{Acknowledgements}  We thank Dima Kharzeev for 
asking the question which stimulated this work.  
This
manuscript has been authorized under Contract No. DE-AC02-98CH10886 with
the U. S. Department of Energy.

\end{document}